# Diffusion broadening of the point spread function in steady-state MRI

Bibek Dhakal[1,2*] and John C. Gore[1,2,3,4]

[1] Vanderbilt University Institute of Imaging Science, Vanderbilt University Medical Center, Nashville, TN, USA

[2] Department of Physics and Astronomy, Vanderbilt University, Nashville, TN, USA

[3] Department of Biomedical Engineering, Vanderbilt University, Nashville, TN, USA

[4] Department of Radiology and Radiological Sciences, Vanderbilt University Medical Center, Nashville, TN, USA

**Correspondence.** Bibek Dhakal (**Email:** bibek.dhakal@vanderbilt.edu)

**Address.** Vanderbilt University Institute of Imaging Science, 1161 21st Avenue South, Nashville, TN 37232, USA.

**Running title.** Flip-angle dependence of diffusive blurring

**Word count.** ~3609 words

## Abstract

**Purpose:** To demonstrate how diffusion blurs the longitudinal magnetization in steady-state gradient-echo imaging, quantify resolution dependence on flip angle and repetition time, and the effects on relaxometry at microscopic resolution.

**Methods:** Diffusion was incorporated into the Bloch-Torrey equation in **k** space to determine how the point spread function (PSF) of $T_1$-weighted spoiled gradient-echo images varies with flip angle. Theoretical predictions were validated numerically and extended to b-SSFP.

**Results:** Diffusion acts as a low-pass filter whose width depends on the diffusion rate ($D$), repetition time ($TR$), $T_1$ and flip angle (α). The blurring length spans from $\sqrt{(D \cdot TR/2)}$ at large α to $\sqrt{(DT_1)}$ as the α→0, equaling $\sqrt{(DT_1/2)}$ at the Ernst angle. Consequently, multi-angle acquisitions have mismatched resolution (PSF widths of 39 μm at α = 10° versus 11 μm at α = 80° for $TR$ = 50 ms). For the flip-angle pair optimizing variable-flip-angle $T_1$ mapping, the effective resolutions differ by a factor of 2.4. In simulated DESPOT1 maps the $R_1$ of a 30 μm feature is reduced to 1.80 $s^{-1}$ (true value 2.00 $s^{-1}$). In b-SSFP blurring depends on both $T_1$, $T_2$, and α but is independent of $TR$, with minimum of $\sqrt{(DT_2)}$.

**Conclusion:** The PSF of a steady-state sequence depends on the flip angle and $T_1$. Combining images acquired at different flip angles implicitly assumes identical spatial resolution, but that is not valid at microscopic scales, affecting the accuracy of quantitative relaxometry. The theory identifies a new limitation on the ultimate resolution of MRI set by diffusion.

## Introduction

The resolution attainable in magnetic resonance microscopy is limited by several well-characterized processes. Signal-to-noise ratio falls as the cube of the voxel dimension, susceptibility differences broaden the intrinsic linewidth, gradient strengths limit the achievable encoding bandwidth, and molecular diffusion in the imaging gradients attenuates high spatial frequencies during spatial encoding. These limits have been well described previously [1,2] and the diffusion effects in particular have been evaluated in detail both analytically and experimentally [3,4]. In the frequency-encoding gradient, diffusion attenuation increases as $exp(-\gamma^2 D G^2 t^3/12)$, so that it becomes particularly severe in the regime of strong gradients ($G$) and long readouts *(t)* required for microscopic resolution [3].

Each of these processes reduces transverse magnetization during acquisitions, but the image is not in fact a map of transverse magnetization. Rather, in a typical fast spoiled gradient-echo acquisition, the measured transverse signal is created by an RF pulse and is proportional to the longitudinal magnetization $M_z$ immediately before it, so that the image is a map of the spatial distribution of the steady state value of $M_z$.

During a sequence of RF excitation pulses, $M_z$ in tissue recovers continuously by spin lattice relaxation toward the local equilibrium value, and is also reduced by RF saturation, while also diffusing by Brownian motion. When such a quantity undergoes diffusion and is reduced at a finite rate, its spatial distribution acquires a characteristic width in the steady state, and structure finer than that length is blurred. This behavior is found in situations such as spin diffusion toward paramagnetic relaxation sinks in solids (Bloembergen [5], Blumberg [6]). The influence of molecular diffusion on relaxation measurements in compartmentalized systems was previously analyzed by Brownstein and Tarr [7]. To our knowledge, however, diffusion of longitudinal magnetization has not previously been identified as a limitation on the imaging point spread function of a steady-state sequence. As shown below, a novel conclusion of note is that the resulting blur depends on the flip angle and $T_1$ and sets a limit on the achievable spatial resolution. Moreover, variable-flip-angle methods for $T_1$ mapping, such as the widely used DESPOT1 [8], combine two or more spoiled gradient-echo images acquired at different flip angles and then invert the signal equation. The inversion assumes that the images differ only in their flip angles. We show below that at microscopic resolution they also differ in their effective resolution, and that this produces a systematic and previously unrecognized bias in the fitted parameters.

Here we derive a transfer function that describes how different spatial frequencies are attenuated by the diffusion of longitudinal magnetization during steady state imaging. From this we derive the characteristic diffusion length that affects longitudinal magnetization for representative acquisitions and quantify the consequences for image resolution and variable-flip-angle relaxometry. The theory is validated using a numerical simulation of a two-point DESPOT1 measurement. The Appendix extends the analysis to balanced steady-state free precession, where the transverse magnetization is not spoiled and both $T_1$ and $T_2$ affect the resolution.

## Theory

In a typical fast gradient echo imaging sequence such as FLASH [9], with gradient spoiling, the transverse magnetization measured at the echo time TE is $M_{xy} = M_z \sin\alpha$, where $M_z$ is the longitudinal magnetization immediately before the excitation pulse. When $\sin\alpha$ is spatially uniform, the image is a map of $M_z$.

In SPGR sequences the transverse magnetization is destroyed at every repetition, so $M_{xy}$ retains no history between repetitions, whereas longitudinal magnetization survives from one repetition to the next, albeit multiplied by $E_1 \cos\alpha$ where $E_1 = exp(-R_1 TR)$. In addition, longitudinal magnetization recovers towards equilibrium. Thus $M_z$ in total may be regarded as the sum of multiple components, each with its own history of RF exposure. The Bloch equations describe the time-dependence of the net longitudinal component,[10]

$$\frac{\partial M_z(\mathbf{r},t)}{\partial t} = R_1[M_0(\mathbf{r}) - M_z(\mathbf{r},t)]$$

which includes spin lattice relaxation toward the local equilibrium value $M_0$. In tissue the magnetization is carried by protons in water molecules which diffuse with self-diffusion coefficient $D$. Torrey [11] added a diffusion term to the Bloch equations:

$$\frac{\partial M_z(\mathbf{r},t)}{\partial t} = R_1[M_0(\mathbf{r}) - M_z(\mathbf{r},t)] + D\nabla^2 M_z(\mathbf{r},t) \qquad (1)$$

$D$ is the quantity measured by diffusion-weighted imaging, and the Laplacian is derived from Fick's second law and says simply that magnetization flows in the direction of a gradient of $M_z$.

The spatial mean $\overline{M}_z$ can be separated from local fluctuations by writing $M_z(\mathbf{r}, t) = \overline{M}_z(t) + \delta M_z(\mathbf{r}, t)$ and similarly $M_0(\mathbf{r}) = \overline{M}_0 + \delta M_0(\mathbf{r})$. The mean obeys the ordinary Bloch equation, the Laplacian of a constant vanishes, leaving an equation for the fluctuations alone, a function of position and time:

$$\frac{\partial\, \delta M_z(\mathbf{r}, t)}{\partial t} = R_1[\delta M_0(\mathbf{r}) - \delta M_z(\mathbf{r}, t)] + D\nabla^2 \delta M_z(\mathbf{r}, t) \qquad (2)$$

The Fourier transform with respect to the spatial coordinate describes the behavior in **k** space

$$\widetilde{\delta M}_z(\mathbf{k}, t) = \int \delta M_z(\mathbf{r}, t)\, e^{-i\mathbf{k}\cdot\mathbf{r}}\, d^3r \qquad \text{and} \qquad \delta M_z(\mathbf{r}, t) = \frac{1}{(2\pi)^3} \int \widetilde{\delta M}_z(\mathbf{k}, t)\, e^{i\mathbf{k}\cdot\mathbf{r}}\, d^3k$$

The Laplacian of $e^{i\mathbf{k}\cdot\mathbf{r}}$ gives a term $-k^2 e^{i\mathbf{k}\cdot\mathbf{r}}$, where $k = |\mathbf{k}|$, so (2) becomes, for each **k** separately,

$$\frac{\partial\, \widetilde{\delta M}_z(\mathbf{k}, t)}{\partial t} = -(R_1 + Dk^2)\widetilde{\delta M}_z(\mathbf{k}, t) + R_1 \widetilde{\delta M}_0(\mathbf{k}) \qquad (3)$$

$Dk^2$ is the rate at which diffusion attenuates magnetization with spatial frequency $k$. For example, the modulation within a sinusoidal variation of period $P = 2\pi/k$ decays with time constant $1/(Dk^2) = P^2/4\pi^2 D$, so contrast from high frequencies decays more rapidly. A modulation with spatial wavelength of 50 microns decays in about 60 ms ($D$ = $10^{-5}$ cm² $s^{-1}$), much faster than $T_1$ in most samples of interest unless they are doped with a paramagnetic agent.

Equation (3) shows there is a direct competition between two rates. Relaxation re-establishes the tissue's magnetization locally at rate $R_1$, as $\widetilde{M}_z$ relaxes toward the local $\widetilde{M}_0(\mathbf{r})$. Diffusion smooths the image of $\widetilde{M}_z$ at rate $Dk^2$, by transferring magnetization between neighbouring regions. For low spatial frequencies relaxation dominates and the intrinsic pattern is faithfully imaged. At fine scales, diffusion may dominate and blurring occurs as high spatial frequencies are attenuated. The two rates are equal at $k = \sqrt{R_1/D}$, which equates typically to a wavelength of ≈200 microns in tissues.

To solve and evaluate equation (3), put $\Lambda = R_1 + Dk^2$ for the total rate at which $\widetilde{\delta M}_z$ decays. Then $\partial\widetilde{\delta M}_z(\mathbf{k},t)/\partial t = -\Lambda\widetilde{\delta M}_z(\mathbf{k},t) + R_1\widetilde{\delta M}_0(\mathbf{k})$. The right-hand side vanishes when

$$\widetilde{\delta M}_z^*(\mathbf{k}) = \frac{R_1\widetilde{\delta M}_0(\mathbf{k})}{R_1 + Dk^2} \qquad (4)$$

$\widetilde{\delta M}_z^*(\mathbf{k})$ is the amplitude of an inhomogeneity that would be reached in equilibrium with no RF pulses. It is not the intrinsic contrast but a reduced fraction of it as relaxation restores the pattern while diffusion simultaneously attenuates it. The two reach an equilibrium short of full amplitude. Equation (4) is a Lorentzian, with characteristic width $\sqrt{D/R_1}$.

If we consider situations (e.g. after some RF disturbance) when the magnetization deviates from this equilibrium, we can define the deviation as $y = \widetilde{\delta M}_z - \widetilde{\delta M}_z^*$ so that, from (4), $dy/dt = -\Lambda y$, i.e. $y$ (the deviation) decays exponentially towards equilibrium as expected. Substituting back,

$$\widetilde{\delta M}_z(k,t) = \widetilde{\delta M}_z^*(k) + \left[\widetilde{\delta M}_z(k,0) - \widetilde{\delta M}_z^*(k)\right]e^{-\Lambda t}$$

which is analogous to the familiar Bloch solution with two differences: the rate is $\Lambda = R_1 + Dk^2$ rather than $R_1$, and the asymptote is $\widetilde{\delta M}_z^*$ rather than the full equilibrium magnetization.

In a steady state acquisition, for one repetition period TR we can define $u = Dk^2TR$. Then $e^{-\Lambda TR} = e^{-R_1TR}e^{-Dk^2TR} = E_1e^{-u} = E$, so

$$\widetilde{\delta M}_z(k,TR) = \widetilde{\delta M}_z^*(k) + \left[\widetilde{\delta M}_z(k,0) - \widetilde{\delta M}_z^*(k)\right]E$$

In a series of RF pulses, each excitation pulse multiplies whatever longitudinal magnetization remains by $\cos\alpha$. Writing $X(k)$ for the amplitude immediately before a pulse, the amplitude just after it is $X(k)\cos\alpha$ and in the steady state

$$X(k) = \widetilde{\delta M}_z^*(k) + \left(X(k)\cos\alpha - \widetilde{\delta M}_z^*(k)\right)E$$

or

$$X(k) = \widetilde{\delta M}_z^*(k) \cdot \frac{1-E}{1-E\cos\alpha}$$

The transfer function for image formation is the ratio of the measured amplitude to that in the object, normalized at $k = 0$. Thus

$$H(k) = \frac{X(k)}{\widetilde{\delta M}_0(k)} \Big/ \frac{X(0)}{\widetilde{\delta M}_0(0)}$$

Writing $q = E_1 \cos\alpha$ so that $E\cos\alpha = qe^{-u}$,

$$H(k) = \frac{R_1}{R_1 + Dk^2} \cdot \frac{1 - E_1 e^{-u}}{1 - E_1} \cdot \frac{1 - q}{1 - qe^{-u}} \qquad (5)$$

Equation (5) is the equivalent of the modulation transfer function for image formation assuming ideal spoiling, uniform flip angle, and uniform $D$ and $R_1$. Figure 1 shows its form for typical values of interest and for two different choices of the flip angle. Table 1 provides values of the diffusion length $\lambda$ defined below and the width of the PSF obtained by inverse Fourier transform of (5). Figure 2 shows the corresponding real-space line-spread function and 3-D radial kernel from the Lorentzian approximation $H(k) = 1/(1 + \lambda^2 k^2)$ with $\lambda$ from Eq. (7). The small flip-angle PSF is broad and low while the large flip-angle PSF is sharp and concentrated near the origin. Thus, the two images used in VFA technique do not share the same spatial resolution.

In (5) $|qe^{-u}| < 1$ so the third factor may be expanded as an infinite geometric series,

$$\frac{1 - q}{1 - qe^{-u}} = \sum_{j=0}^{\infty} w_j \, e^{-uj}, \qquad w_j = (1 - q)q^{\,j} \qquad (6)$$

The measured magnetization in the steady state can thus be viewed as a superposition of contributions of different ages or histories that have “seen” different numbers of RF pulses. The fraction $w_j$ is created during the repetition that ended $j$ pulses earlier so the factor $e^{-uj}$ represents the diffusion effects accumulated over $j$ pulse intervals. The weights form a geometric distribution with mean $\langle j \rangle = q/(1 - q)$. The image is a time-averaged representation of diffusing components. The flip angle determines the relevant averaging time through $q$. A near-90° pulse zeroes $M_z$ and relaxation partly re-establishes it during the next repetition, which depends only on the local $M_0$ and $R_1$. On the other hand, a small-flip-angle pulse barely disturbs $M_z$ and the measured magnetization is derived from components created over many repetitions that have had a long time to diffuse. Figure 3 plots the values of $w_j e^{-uj}$ for different $j$ for two acquisitions with different flip angles.

The first two factors of (5) together describe the blurring produced in a single $TR$ of freshly created magnetization. Their product does not depend on flip angle. For studies which involve comparisons between images with different flip angles, it is common to each image, and all of the flip-angle dependence of the resolution resides in the third factor.

## The diffusion length

Expanding (5) to first order in $u$ gives a characteristic length. The third factor contributes $1 - u\langle j \rangle$. In the first two factors the Lorentzian $R_1/(R_1 + Dk^2) \approx 1 - Dk^2/R_1$ is offset by $(1 - E_1 e^{-u})/(1 - E_1) \approx 1 + uE_1/(1 - E_1)$, and when $R_1 TR \ll 1$ the two terms in $Dk^2/R_1$ cancel exactly, leaving $1 - u/2$. This may be interpreted as the observation that freshly created magnetization has diffused for a mean time of half a repetition. Hence $H(k) \approx 1 - k^2\lambda^2$ with

$$\lambda^2 = D\,TR\left(\frac{1}{2} + \frac{q}{1 - q}\right) = D\langle t \rangle, \qquad \langle t \rangle = \left(\langle j \rangle + \frac{1}{2}\right)TR \qquad (7)$$

$\lambda$ is the distance water diffuses during the mean age of the measured magnetization. $\langle j \rangle$ is set by $q$, so $\lambda$ is bounded above and below:

$$\sqrt{D\,TR/2} \;\le\; \lambda \;\le\; \sqrt{D\,T_1} \qquad (8)$$

The lower limit is approached as $\alpha \to 90°$, when the measured magnetization is created entirely within the last repetition. The upper limit is reached as $\alpha \to 0$, when RF saturation is negligible and the averaging time is the full $T_1$. At the Ernst angle, where $q = E_1^2$, (7) gives $\lambda = \sqrt{D\,T_1/2}$ independent of $TR$.

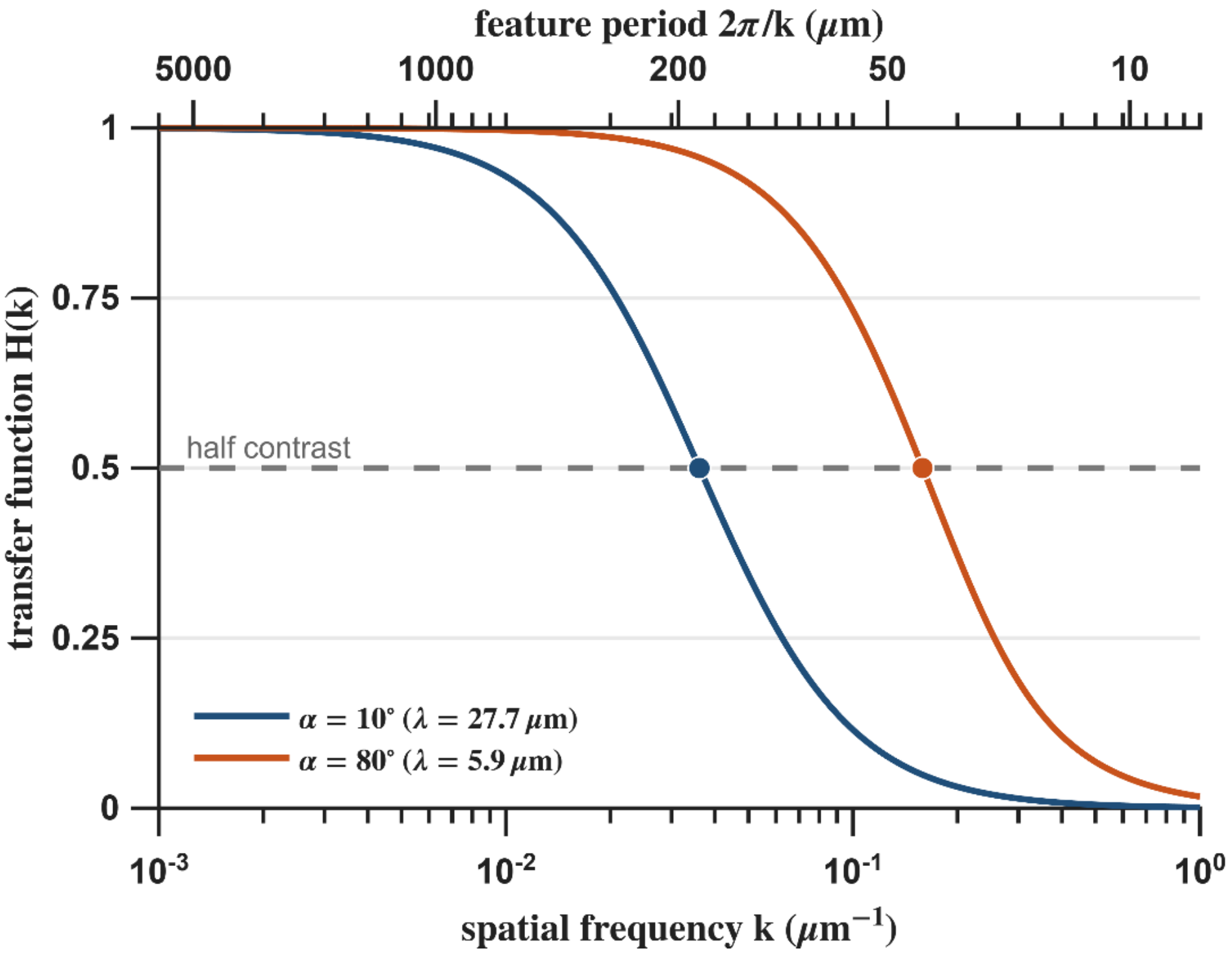

**Figure 1** Calculated $H(k)$ for the SPGR steady state, Eq. (5), with $R_1$ = 1 s$^{-1}$, $D$ = 10$^{-5}$ cm² s$^{-1}$ and $TR$ = 50 ms. Markers show the half-contrast frequency. Values of $\lambda$ are from Eq. (7).

**TABLE 1** Diffusion lengths and PSF width for a spoiled gradient-echo steady state. Values assume $R_1$ = 1 s$^{-1}$ and $D$ = 10$^{-5}$ cm² s$^{-1}$ (= 1 μm² ms$^{-1}$). $\lambda$ is from Eq. (7) and scales as $\sqrt{D}$; the PSF width is the full width at half maximum of the inverse Fourier transform of Eq. (5).

| $TR\ (ms)$ | $\alpha$ | $\lambda\ (\mu m)$ | PSF FWHM ($\mu m$) |
|---|---|---|---|
| 25 | 10° | 24.9 | 35 |
| 25 | 30° | 12.2 | 18 |
| 25 | 80° | 4.2 | 8 |
| 50 | 10° | 27.7 | 39 |
| 50 | 30° | 16.1 | 23 |
| 50 | 80° | 5.9 | 11 |

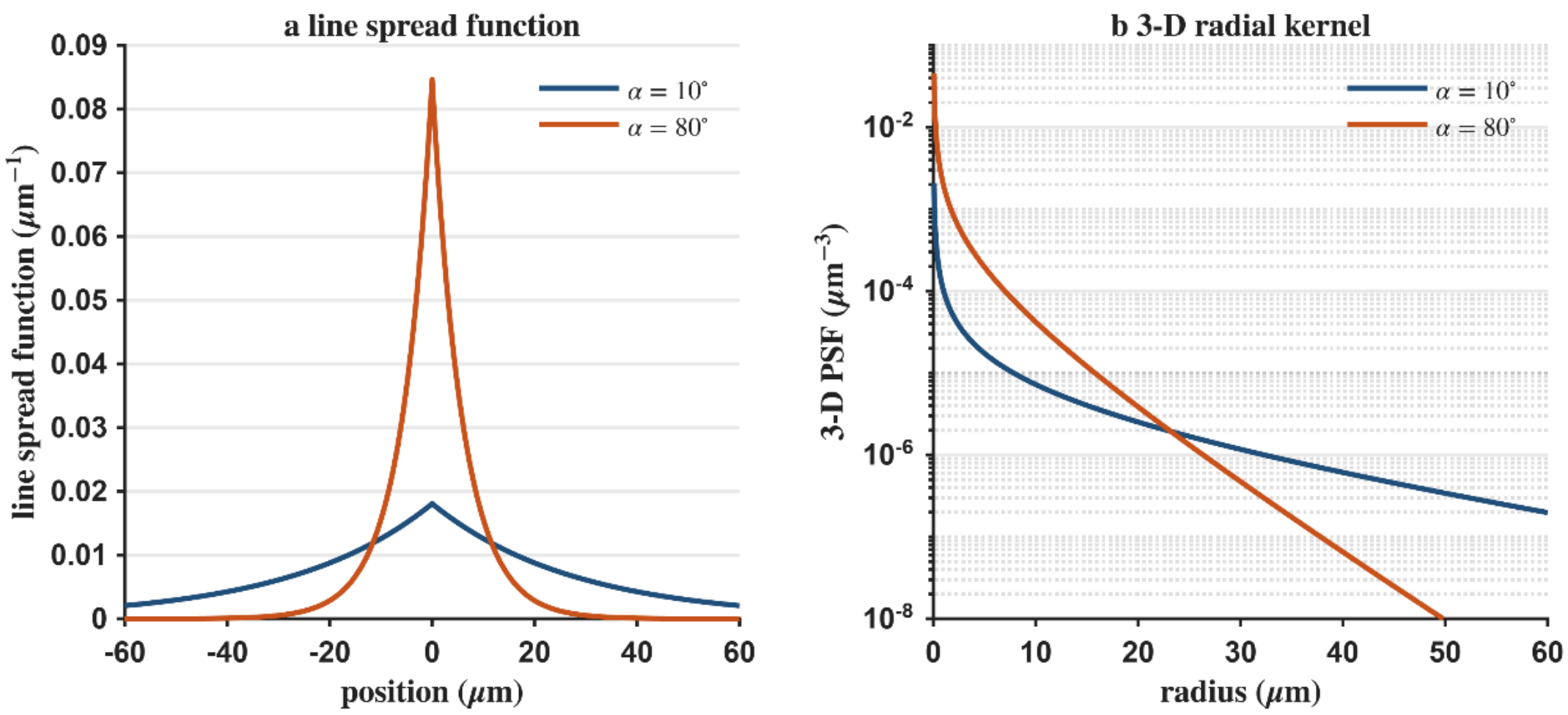

**Figure 2** Line-spread function and 3-D radial point-spread kernel from the Lorentzian approximation $H(k) = 1/(1 + \lambda^2 k^2)$ with $\lambda$ from Eq. (7), for the same parameters as Figure 1 ($R_1$ = 1 s$^{-1}$, $D$ = 10$^{-5}$ cm² s$^{-1}$ and $TR$ = 50 ms). (**A**) One-dimensional line spread $\frac{exp\left(-\frac{|x|}{\lambda}\right)}{2\lambda}$. (**B**) 3D radial kernel $\frac{exp\left(-\frac{r}{\lambda}\right)}{4\pi\lambda^2 r}$. At $\alpha$ = $10^{\circ}(\lambda = 27.7\ \mu\text{m})$ the PSF is broad while at $\alpha$ = $80^{\circ}(\lambda = 5.9\ \mu\text{m})$, it is narrow.

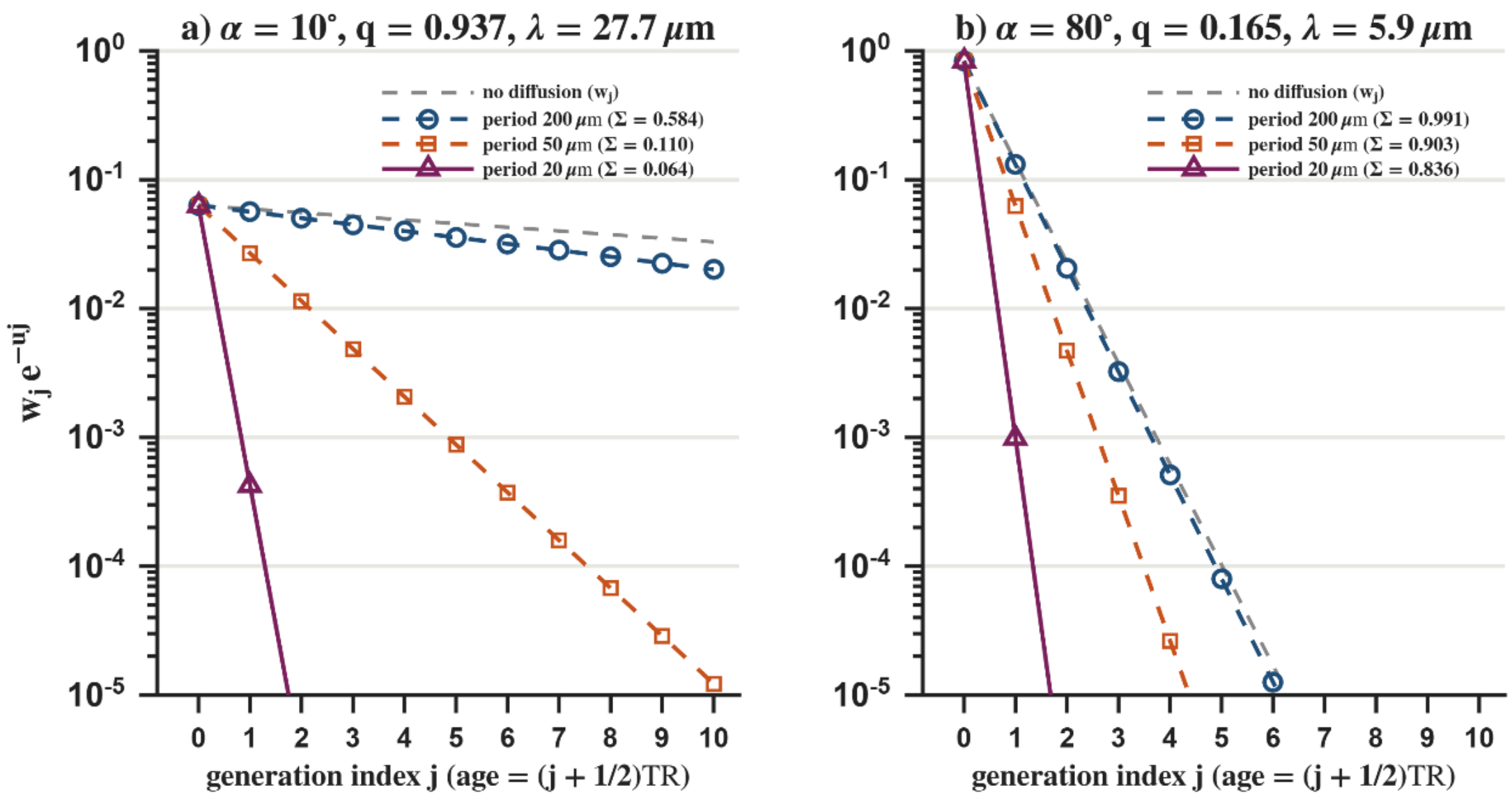

**Figure 3** Contributions $w_j e^{-uj}$ to the third factor of (5) as a function of generation index $j$, for feature periods of 200, 50 and 20 μm, at (**A**) 10° and (**B**) 80°. The dashed line is $w_j$ alone, the no-diffusion limit. $\Sigma$ is the sum over all $j$. At 80° the measurement is dominated by magnetization created within the current repetition, which no amount of diffusion in earlier repetitions can blur; at 10° the weight extends over many repetitions and fine structure is lost. $R_1$ = 1 s$^{-1}$, $D$ = 10$^{-5}$ cm² s$^{-1}$, $TR$ = 50 ms.

# Methods

## Numerical Simulation

The transfer function of Equation (5) is a linear response derived for uniform $R_1$. To test it, and to quantify the consequences for variable-flip-angle relaxometry, the steady state of the Bloch-Torrey equation was solved for an object in which $R_1$ and $M_0$ both vary. A one-dimensional periodic grid of 1600 points at 0.5 μm spacing was used. A single feature with $R_1$ = 2.0 s$^{-1}$ and $M_0$ = 0.70 was placed in a matrix with $R_1$ = 1.0 s$^{-1}$ and $M_0$ = 0.90, with $D$ = 10$^{-5}$ cm$^2$ s$^{-1}$ uniform throughout. Diffusion and relaxation over each TR were applied with the matrix exponential of the discrete periodic Laplacian. Steady-state images were formed at flip angles of 10° and 80° with $TR$ = 50 ms, both with diffusion and, for comparison, with $D$ = 0. Feature widths of 30 $\mu$m and 200 $\mu$m were used for the maps in Figure 4. The width was then varied from 10 to 300 $\mu$m with all other parameters fixed. $R_1$ was estimated voxel by voxel using the standard two-point DESPOT1 equation. All simulations and DESPOT1 fits were performed in MATLAB R2026a (The MathWorks, Natick, MA).

# Results

Figure 4 shows the flip-angle images and the resulting DESPOT1 $R_1$ maps for 30 μm and 200 μm features. At 10° the signal is strongly blurred relative to the $D$ = 0 reference, whereas at 80° the profile remains much sharper. Combining these images in DESPOT1 fitting yields a center $R_1$ of only 1.80 s$^{-1}$ for the 30 μm feature against a true value of 2.00 s$^{-1}$, so a fifth of the true contrast is lost, and the fitted $M_0$ is 0.752 against a true 0.700. For the 200 μm feature the center is recovered as 1.998 s$^{-1}$ and $M_0$ as 0.701. The edges are smeared over a 10–90% width of 21 μm for the 30 μm feature and 28 μm for the 200 μm feature, and the elevation of $R_1$ in the area outside the feature falls to a tenth of the true contrast about 14 μm beyond the boundary.

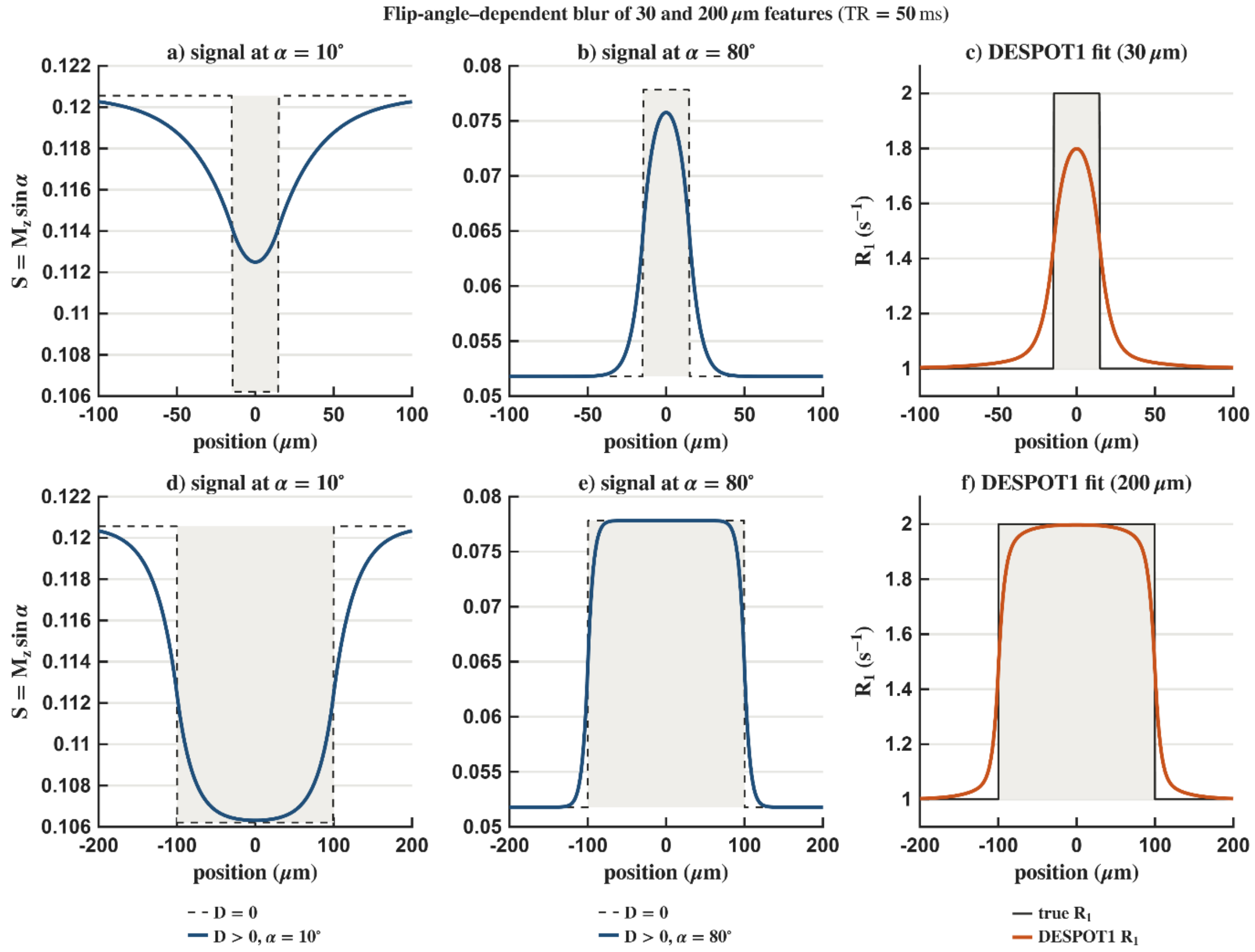

**Figure 4** Flip-angle dependent diffusion blur of a strip feature and the two-point DESPOT1 $R_1$ fit. A feature with $R_1$ = 2.0 s$^{-1}$ and $M_0$ = 0.70 is placed in a matrix with $R_1$ = 1.0 s$^{-1}$ and $M_0$ = 0.90 ($D$ =10$^{-5}$ cm² s$^{-1}$ and $TR$ = 50 ms) Top row (**A-C**): 30 µm feature and bottom row (**D-F**): 200 µm feature. Shading marks true extent of the feature. (**A,D**) Transverse signal at flip angle 10°, (**B**,**E**) 80°, (**C,F**) DESPOT1 $R_1$ from the two flip-angle images (orange) versus the true $R_1$ (black). At 30 µm the fitted center is 1.80 s$^{-1}$ (true 2.00 s$^{-1}$) and at 200 µm the center is recovered except within about 20 µm of its edges.

Figure 5 shows how this bias depends on feature size. The fitted center $R_1$ rises from 1.42 $s^{-1}$ at 10 $\mu$m (-29% bias) towards the true value of 2.00 $s^{-1}$. At 30 $\mu$m it is 1.80 $s^{-1}$ (-10%), matching Figure 4, and by 200 $\mu$m the bias is below 0.2%. The transition occurs around the diffusion lengths of the two acquisitions (shaded region in Figure 5), confirming that DESPOT1 error is severe when the object structure is on the scale of $\lambda$ and becomes negligible once the feature is much wider than $\lambda$ (10°).

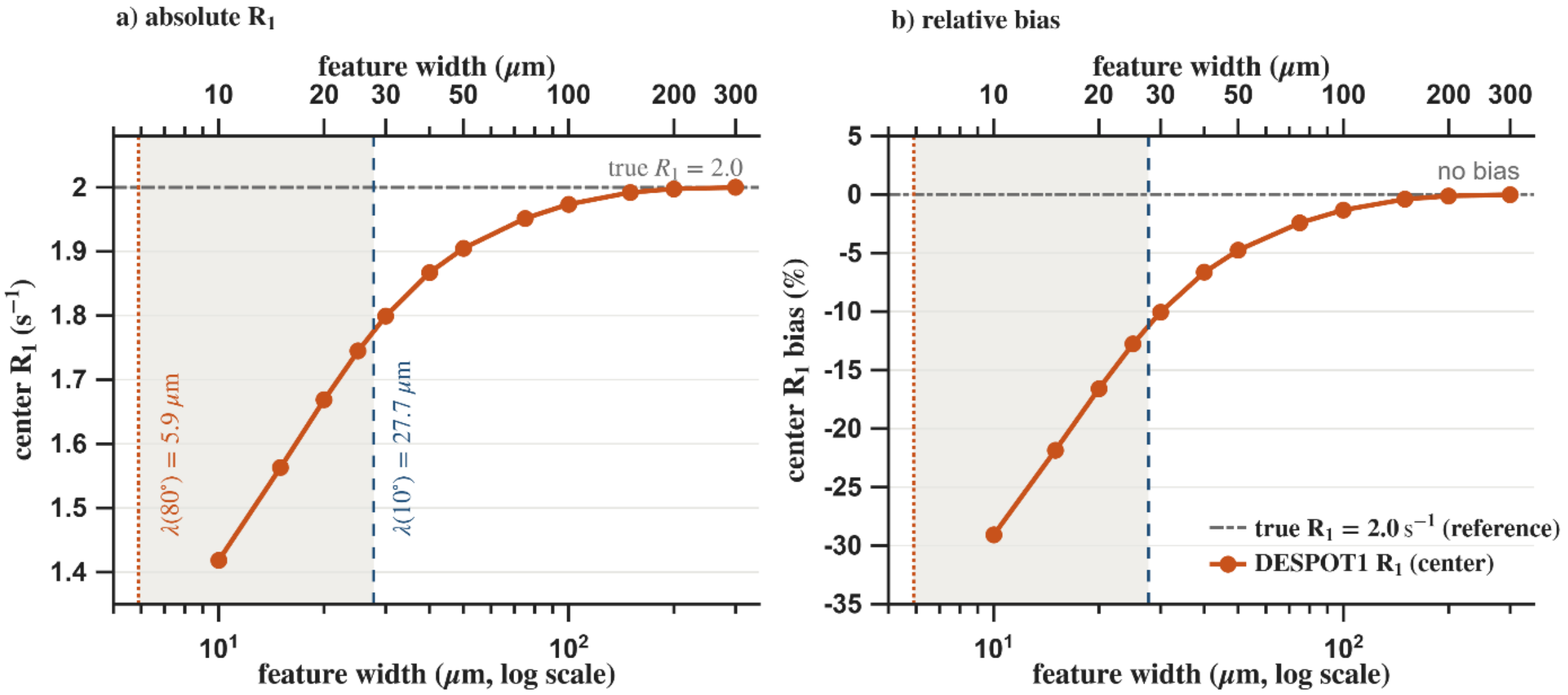

**Figure 5** DESPOT1 bias versus feature width. The same object as in Figure 4 (feature $R_1$ = 2.0 s⁻¹, $M_0$ = 0.70; matrix $R_1$ = 1.0 s⁻¹, $M_0$ = 0.90; $D$ = $10^{-5}$ cm² s⁻¹; $TR$ = 50 ms; flip angles 10° and 80°) was simulated for feature widths from 10 to 300 μm. (**A**) Fitted center $R_1$, (**B**) relative bias. Vertical lines mark the diffusion lengths $\lambda$(80°) = 5.9 μm and $\lambda$(10°) = 27.7 μm from Equation (7) in the matrix and the shaded band lies between them. Bias is large when the feature is comparable to $\lambda$ and falls below about 1% only for widths $\gtrsim$ 150 μm.

## Discussion

The effects described above are intrinsic to the sequence and are present in any steady-state acquisition. At conventional image resolution the effect is small. However, it becomes important when the diffusion lengths approach the voxel dimension, that is, below roughly 50 µm, which is the regime of MR microscopy [1,2,12-14].

Of note, (8) shows that the blurring cannot be avoided by lowering the flip angle: as saturation is removed the averaging time lengthens to $T_1$ and $\lambda$ rises to $\sqrt{D\,T_1}$, which is 22 µm for tissue with $T_1$ ≈ 500 ms and $D$ = 1 µm² ms$^{-1}$, and ≈32 µm for $T_1$ ≈ 1000 ms. Sharper images of $M_z$ require stronger saturation, and the sharpest attainable is $\sqrt{D\,TR/2}$ at $\alpha \to 90°$. The point spread function has unit area as $H(0) = 1$ so the mean of $M_z$ over an inhomogeneous region containing is preserved, and only the contrast between neighbouring features is attenuated.

**Implications for quantitative relaxometry.** Variable flip-angle $T_1$ mapping has been analyzed previously for several sources of systematic error, including transmit field inhomogeneity [15,16], imperfect spoiling [17,18], and magnetization transfer [19]. The present effect is distinct from all of these and is not removed by any of the corresponding corrections. It is a difference in the effective resolution of the images being combined, as illustrated by the 10° and 80° profiles in Figure 2, 4. It biases fitted parameters wherever the object contains structure on the scale of the diffusion length.

The mismatch is intrinsic to efficient measurements. Taking the pair of flip angles that maximizes precision in DESPOT1, at which the signal is 0.71 of its maximum value, (7) gives $\lambda = 0.92\sqrt{D\,T_1}$ for the lower angle and $0.385\sqrt{D\,T_1}$ for the higher, a ratio of 2.40 that is independent of $D$, $T_1$ and $TR$. An optimally designed variable-flip-angle experiment therefore always combines two images whose effective resolutions differ by a factor of about 2.4. The same kind of mismatch arises for variable-$TR$ $R_1$ methods when a fixed flip angle is used, since $\lambda$ also depends on $TR$ through Equation (7), so images acquired with different repetition times have different effective resolutions and cannot be combined without bias at microscopic scales. However, this route is already unattractive for MR Microscopy, since multi-$TR$ acquisitions are much less SNR-efficient than variable-flip-angle approaches.

These conclusions apply to sequences in which the transverse magnetization is spoiled. The Appendix uses a similar analysis of balanced steady-state free precession, in which magnetization is not destroyed between repetitions. The blurring length is then dependent on $R^* = R_1\cos^2(\alpha/2) + R_2\sin^2(\alpha/2)$, is independent of $TR$ once $TR \ll T_2$, and cannot be reduced below $\sqrt{D\,T_2}$. At the flip angles and repetition times at which balanced SSFP is normally used, the diffusion length is several times larger than that of a spoiled sequence with the same timing.

The competition between diffusion and relaxation is long established in other settings. In imaging, diffusion has been treated as a source of edge effects near boundaries in the presence of gradients [20,21] and as an attenuation of high spatial frequencies during encoding [3,4,22]. However, the effect of diffusion on longitudinal magnetization in a steady-state sequence, the resulting dependence of the effective point spread function on flip angle, and the consequence for methods that combine images acquired at different flip angles, have not been previously explained.

The theory above assumes ideal spoiling [23], so that the pulse acts on $M_z$ as a pure multiplication by $\cos\alpha$; with imperfect spoiling, stimulated-echo pathways return transverse coherence to the longitudinal axis and the recursion acquires further terms. It assumes a spatially uniform flip angle. Diffusion is taken to be uniform, scalar and unrestricted; in tissue the effective diffusivity is properly a tensor, which predicts anisotropic blurring that has not been examined here. In white matter the point spread function would then be elongated along the fibre direction, with $\lambda$ differing by the square root of the ratio of the principal diffusivities, a factor of roughly 1.7 for a typical axial-to-radial ratio of 3.

In a steady-state gradient-echo experiment the image is a map of the longitudinal magnetization, and that magnetization undergoes diffusion and is altered by both relaxation and RF saturation. The measured signal typically comprises multiple components with different RF and diffusion histories. The point spread function then depends on the flip angle as well as other parameters. Methods that combine images acquired at different flip angles therefore combine images of different effective resolution, and at microscopic scales the resulting bias can be substantial. The effect is absent in homogeneous media, and is distinct from diffusive attenuation in the imaging gradients. The same analysis applied to balanced steady-state free precession shows that the blurring there is set by both relaxation times and the flip angle, and cannot be reduced by shortening $TR$.

## Appendix: Extension to Steady State Free Precession

In a spoiled sequence only $M_z$ evolves during a series of RF pulses. In balanced SSFP no magnetization is destroyed by the RF train, and instead the pulses rotate magnetization between the longitudinal and transverse axes and the effects of the encoding gradients are refocused within each repetition. We can analyze the effects of on-resonance excitation pulses of alternating phase, ideally balanced gradients, and uniform $R_1$, $R_2$ and $D$.

Torrey's diffusion term acts on each Cartesian component. The longitudinal equation is Equation (1) unchanged, and its transverse counterpart is

$$\frac{\partial M_x}{\partial t} = -R_2 M_x + D\nabla^2 M_x, \qquad \frac{\partial M_y}{\partial t} = -R_2 M_y + D\nabla^2 M_y \qquad (A1)$$

Separating the means from the fluctuations and transforming with respect to position, exactly as in Equations (2) and (3), each mode spatial frequency **k** evolves and diffusion appears as an addition of $Dk^2$ to both relaxation rates. The longitudinal component obeys Equation (3) and the transverse components decay at $R_2 + Dk^2$. Integrating over one repetition period and writing $E_2 = e^{-R_2 TR}$,

$$E = E_1 e^{-u}, \qquad F = E_2 e^{-u}, \qquad \widetilde{\delta M}_\perp(TR) = \widetilde{\delta M}_\perp(0)\, F \qquad (A2)$$

with the longitudinal component relaxing by the factor *E*, as in the main text. Each spatial frequency therefore undergoes, over one repetition, exactly the evolution of a conventional balanced SSFP voxel, with three substitutions:

$$E_1 \to E_1 e^{-u}, \qquad E_2 \to E_2 e^{-u}, \qquad M_0(1 - E_1) \to \widetilde{\delta M}_z^*(1 - E) \qquad (A3)$$

The first two state that each relaxation rate is augmented by the rate at which diffusion erases a modulation of wavenumber *k*; the third states that the source of new magnetization is itself reduced, because the pattern relaxation restores is the partially reduced one of Equation (4).

The steady state is that of the standard on-resonance balanced SSFP sequence [24,25] with the substitutions (A3) applied. Immediately after the excitation pulse,

$$X_{\perp}(k) = \widetilde{\delta M_z^*}(k)\ \frac{(1-E)\sin\alpha}{1-(E-F)\cos\alpha - EF} \qquad (A4)$$

Normalizing to unity at $k$ = 0 as before gives the transfer function

$$H_{\text{bSSFP}}(k) = \frac{R_1}{R_1 + Dk^2} \cdot \frac{1-E_1 e^{-u}}{1-E_1} \cdot \frac{1-(E_1-E_2)\cos\alpha - E_1E_2}{1-(E_1-E_2)\cos\alpha\, e^{-u} - E_1E_2\, e^{-2u}} \qquad (A5)$$

which is multiplied by a further factor $e^{-u/2}$ if the signal is sampled at $TE = TR/2$ rather than immediately after the pulse. The first two factors are identical to those of Equation (5): the contrast created by $T_1$ recovery within a single repetition does not depend on the flip angle or any refocusing of the transverse magnetization. The difference between the two sequences lies in the third factor, whose denominator is linear in $e^{-u}$ for a spoiled sequence and quadratic. $R_2$ enters only through this factor.

Writing $z = e^{-u}$, $a = (E_1 - E_2)\cos\alpha$ and $b = E_1E_2$, that denominator is the quadratic $1 - az - bz^2$, which factorizes as $(1-\mu_+ z)(1-\mu_- z)$ with

$$\mu_{\pm} = \frac{a \pm \sqrt{a^2+4b}}{2} \qquad (A6)$$

Expanding in powers of $z$ again expresses the measurement as a superposition of contributions created in earlier repetitions, as in Equation (6), but with weights that are the sum of two geometric sequences rather than one,

$$w_j = (1-a-b)\frac{\mu_+^{j+1} - \mu_-^{j+1}}{\mu_+ - \mu_-} \qquad (A7)$$

The root $\mu_+$ is positive and is the counterpart of *q*; for $TR \ll T_2$ it reduces to $\mu_+ \approx 1 - R^*TR$ with $R^* = R_1\cos^2(\alpha/2) + R_2\sin^2(\alpha/2)$, the rate at which a balanced SSFP sequence approaches its steady state [26]. The second root is negative, because the pulses alternate in phase, so contributions created in alternate repetitions have opposite signs.

Expanding (A5) to first order in *u* gives the diffusion length in the same way as Equations (7) and (8). The first two factors again contribute $1 - u/2$, and the third contributes $1 - u\langle j\rangle$ with

$$\langle j\rangle = \frac{(E_1 - E_2)\cos\alpha + 2E_1E_2}{1 - (E_1 - E_2)\cos\alpha - E_1E_2}, \qquad \lambda^2 = D\,TR\left(\frac{1}{2} + \langle j\rangle\right) \qquad (A8)$$

so that $\lambda$ keeps its meaning as the distance water diffuses during the mean age of the measured magnetization. What changes is the age. In a spoiled sequence it is limited by the pulses, which remove a fraction $1 - cos\alpha$ of $M_z$ at every repetition; here nothing is removed and the age is limited only by relaxation, apportioned between $T_1$ and $T_2$ by the flip angle. For $TR \ll T_2$, (A8) reduces to

$$\lambda^2 \to \frac{D}{R^*} = \frac{D}{R_1\cos^2(\alpha/2) + R_2\sin^2(\alpha/2)} \qquad (A9)$$

in which effects of $TR$ are cancelled. In a spoiled sequence $\lambda$ can be reduced without limit by shortening $TR$, the lowest being $\sqrt{D\,TR/2}$ from Equation (8); in balanced SSFP shortening TR does not sharpen the image at all once $TR \ll T_2$, and the lowest limit is $\sqrt{D\,T_2}$, approached as $\alpha \to$ 180°. The ceiling $\sqrt{D\,T_1}$ as $\alpha \to$ 0 is common to both sequences.

**TABLE A1** Diffusion length in balanced SSFP for $T_1$ = 1 s, $T_2$ = 50 ms and $D$ = $10^{-5}$ $cm^2$ $s^{-1}$. $\langle j\rangle$ is the mean number of repetitions of accumulated diffusion from Equation (A8). The last column gives $\lambda$ for a spoiled sequence at the same $TR$ and flip angle, from Equation (7).

| $\boldsymbol{TR}$ ($\boldsymbol{ms}$) | $\boldsymbol{\alpha}$ | $\boldsymbol{\langle j\rangle}$ | $\boldsymbol{\lambda}$ (μ$\boldsymbol{m}$) | $\boldsymbol{\lambda, SPGR}$ (μ$\boldsymbol{m}$) |
|---|---|---|---|---|
| 5 | 10° | 173.8 | 29.5 | 15.7 |
| 5 | 30° | 87.1 | 20.9 | 5.8 |
| 5 | 60° | 33.8 | 13.1 | 2.7 |
| 5 | 80° | 21.6 | 10.5 | 1.9 |
| 50 | 10° | 16.9 | 29.5 | 27.7 |
| 50 | 30° | 8.3 | 21.0 | 16.1 |
| 50 | 60° | 2.8 | 12.8 | 8.4 |
| 50 | 80° | 1.5 | 9.9 | 5.9 |

Only diffusion of magnetization is included above: the imaging gradients also dephase and rephase the transverse magnetization within each repetition, and diffusion during those lobes attenuates the signal. Finally, because $\lambda$ depends on flip angle through $R^*$, methods that fit relaxation times from balanced SSFP images acquired at several flip angles [27], combine images of different effective resolution in the same way as variable-flip-angle $T_1$ mapping.

## Data availability statement

The MATLAB script used for the numerical simulation in this study is available at https://github.com/BibekDhakal977/diffusion-psf-SSMRI.git

## Figure Captions

**Figure 1** Calculated $H(k)$ for the SPGR steady state, Eq. (5), with $R_1$ = 1 s$^{-1}$, $D$ = 10$^{-5}$ cm² s$^{-1}$ and $TR$ = 50 ms. Markers show the half-contrast frequency. Values of $\lambda$ are from Eq. (7).

**Figure 2** Line-spread function and 3-D radial point-spread kernel from the Lorentzian approximation $H(k) = 1/(1+\lambda^2 k^2)$ with $\lambda$ from Eq. (7), for the same parameters as Figure 1 ($R_1$ = 1 s$^{-1}$, $D$ = 10$^{-5}$ cm² s$^{-1}$ and $TR$ = 50 ms). (**A**) One-dimensional line spread $\frac{exp\left(-\frac{|x|}{\lambda}\right)}{2\lambda}$. (**B**) 3D radial kernel $\frac{exp\left(-\frac{r}{\lambda}\right)}{4\pi\lambda^2 r}$. At $\alpha$ = $10^{\circ}(\lambda = 27.7\ \mu\text{m})$ the PSF is broad while at $\alpha$ = $80^{\circ}(\lambda = 5.9\ \mu\text{m})$, it is narrow.

**Figure 3** Contributions $w_j e^{-uj}$ to the third factor of (5) as a function of generation index $j$, for feature periods of 200, 50 and 20 µm, at (**A**) 10° and (**B**) 80°. The dashed line is $w_j$ alone, the no-diffusion limit. $\Sigma$ is the sum over all $j$. At 80° the measurement is dominated by magnetization created within the current repetition, which no amount of diffusion in earlier repetitions can blur; at 10° the weight extends over many repetitions and fine structure is lost. $R_1$ = 1 s$^{-1}$, $D$ = 10$^{-5}$ cm² s$^{-1}$, $TR$ = 50 ms.

**Figure 4** Flip-angle dependent diffusion blur of a strip features and the two-point DESPOT1 $R_1$ fit. A feature with $R_1$ = 2.0 s$^{-1}$ and $M_0$ = 0.70 is placed in a matrix with $R_1$ = 1.0 s$^{-1}$ and $M_0$ = 0.90 ($D$ =10$^{-5}$ cm² s$^{-1}$ and $TR$ = 50 ms) Top row (**A-C**): 30 µm feature and bottom row (**D-F**): 200 µm feature. Shading marks true extent of the feature. (**A,B**) Transverse signal at flip angle 10°, (**C,F**) DESPOT1 $R_1$ from the two flip-angle images (orange) versus the true $R_1$ (black). At 30 µm the fitted center is 1.80 s$^{-1}$ (true 2.00 s$^{-1}$) and at 200 µm the center is recovered except within about 20 µm of its edges.

**Figure 5** DESPOT1 bias versus feature width. The same object as in Figure 4 (feature $R_1$ = 2.0 s$^{-1}$, $M_0$ = 0.70; matrix $R_1$ = 1.0 s$^{-1}$, $M_0$ = 0.90; $D$ = 10$^{-5}$ cm² s$^{-1}$; $TR$ = 50 ms; flip angles 10° and 80°) was simulated for feature widths from 10 to 300 µm. (**A**) Fitted center

$R_1$, (**B**) relative bias. Vertical lines mark the diffusion lengths $\lambda(80°)$ = 5.9 μm and $\lambda(10°)$ = 27.7 μm from Equation (7) in the matrix and the shaded band lies between them. Bias is large when the feature is comparable to $\lambda$ and falls below about 1% only for widths $\gtrsim$ 150 μm.

# Tables

**TABLE 1** Diffusion lengths and PSF width for a spoiled gradient-echo steady state. Values assume $R_1$ = 1 $s^{-1}$ and $D$ = $10^{-5}$ $cm^2$ $s^{-1}$ (= 1 $\mu m^2$ $ms^{-1}$). $\lambda$ is from Eq. (7) and scales as $\sqrt{D}$; the PSF width is the full width at half maximum of the inverse Fourier transform of Eq. (5).

| $TR$ ($ms$) | $\alpha$ | $\lambda$ ($\mu m$) | PSF FWHM ($\mu m$) |
|---|---|---|---|
| 25 | 10° | 24.9 | 35 |
| 25 | 30° | 12.2 | 18 |
| 25 | 80° | 4.2 | 8 |
| 50 | 10° | 27.7 | 39 |
| 50 | 30° | 16.1 | 23 |
| 50 | 80° | 5.9 | 11 |

**TABLE A1** Diffusion length in balanced SSFP for $T_1$ = 1 s, $T_2$ = 50 ms and $D$ = $10^{-5}$ cm² $s^{-1}$. $\langle j \rangle$ is the mean number of repetitions of accumulated diffusion from Equation (A8). The last column gives $\lambda$ for a spoiled sequence at the same $TR$ and flip angle, from Equation (7).

| $TR$ $(ms)$ | $\alpha$ | $\langle j \rangle$ | $\lambda$ (μ$m$) | $\lambda, SPGR$ (μ$m$) |
|---|---|---|---|---|
| 5 | 10° | 173.8 | 29.5 | 15.7 |
| 5 | 30° | 87.1 | 20.9 | 5.8 |
| 5 | 60° | 33.8 | 13.1 | 2.7 |
| 5 | 80° | 21.6 | 10.5 | 1.9 |
| 50 | 10° | 16.9 | 29.5 | 27.7 |
| 50 | 30° | 8.3 | 21.0 | 16.1 |
| 50 | 60° | 2.8 | 12.8 | 8.4 |
| 50 | 80° | 1.5 | 9.9 | 5.9 |